\documentclass[%
 reprint,
 amsmath,amssymb,
 aps,
 nofootinbib,
]{revtex4-2}
\usepackage{enumitem}
\usepackage{hyperref}
\usepackage{xcolor}
\usepackage{braket}
\usepackage{subcaption}
\usepackage{pdfcomment}
\usepackage{todonotes}
\usepackage{nicematrix}
\usepackage{siunitx}[=v2]
\usepackage{graphicx} 
\usepackage[nolist,nohyperlinks]{acronym}
\usepackage{amsmath, amsthm, amssymb, amsbsy, mathtools, mathalpha}

\usepackage{graphicx}% Include figure files
\usepackage{dcolumn}% Align table columns on decimal point
\usepackage{bm}% bold math
\usepackage{hyperref}% add hypertext capabilities
\usepackage{color, units}
\usepackage[normalem]{ulem}

\usepackage{xspace}
\usepackage{comment}
\usepackage{xcolor}

\graphicspath{{figures/}}

\renewcommand{\d}{{\bf d}}
\newcommand{\bnu}{\mbox{\boldmath $\nu$}}
\newcommand{\btheta}{\mbox{\boldmath $\theta$}}
\newcommand{\bdelta}{\mbox{\boldmath $\delta$}}

\renewcommand{\S}{{\bf S}}
\newcommand{\A}{{\bf A}}

\newcommand{\Z}{{\bf Z}}

\newcommand{\eqq}{\ensuremath{\!\mathrel{\scalebox{0.7}{=}}\!}}
\newcommand{\tmes}{\ensuremath{{\mkern-2mu\times\mkern-2mu}}}
\newcommand{\ex}[1]{\ensuremath{\tmes10^{\text{-}#1}}}

\newcommand{\project}[1]{\textsf{#1}}

\newcommand{\python}{\project{Python}}
\newcommand{\scipy}{\project{scipy}}
\newcommand{\jupyterbook}{\project{Jupuyter Book}}
\newcommand{\numpy}{\project{numpy}}
\newcommand{\pandas}{\project{pandas}}
\newcommand{\matplotlib}{\project{matplotlib}}
\newcommand{\hyperopt}{\project{HyperOpt}}
\newcommand{\tensorflowProb}{\project{TensorFlow Probability}}

\DeclareSIUnit{\kilo}{k}

\begin{document}

\preprint{APS/123-QED}

\title{Variational inference for correlated\\ gravitational wave detector network noise
}
\newcommand{\UoAStats}{Department of Statistics, University of Auckland, 38 Princes St, Auckland, New Zealand}
\newcommand{\UAntwerpen}{Universiteit Antwerpen, Prinsstraat 13, 2000 Antwerpen, Belgium}
\newcommand{\UCoteAzur}{Universit\'e C\^ote d'Azur, Observatoire C\^ote d'Azur, CNRS, Artemis, 06304 Nice, France}
\newcommand{\AdelaidePhysics}{Department of Physics, The University of Adelaide, Adelaide, SA 5005, Australia}
\newcommand{\ARCDarkMatter}{ARC Centre of Excellence for Dark Matter Particle Physics, Australia}
\newcommand{\AUTMaths}{Department of Mathematical Sciences, Auckland University of Technology, Auckland, New Zealand}
\newcommand{\CambridgeBioStats}{MRC Biostatistics Unit, University of Cambridge, Cambridge, UK}

\author{Jianan Liu$^{1}$}
\author{Avi Vajpeyi$^{1}$}
\author{Renate Meyer$^{1}$}
\author{Kamiel Janssens$^{2,3,4,5}$}%
\author{Jeung Eun Lee$^{1}$}
\author{Patricio Maturana-Russel$^{1,6}$}
\author{Nelson Christensen$^{3}$}
\author{Yixuan Liu$^{7}$}

\affiliation{$^{1}$\UoAStats}%
\affiliation{$^{2}$\UAntwerpen}
\affiliation{$^{3}$\UCoteAzur}
\affiliation{$^{4}$\AdelaidePhysics}
\affiliation{$^{5}$\ARCDarkMatter}
\affiliation{$^{6}$\AUTMaths}
\affiliation{$^{7}$\CambridgeBioStats}
%\collaboration{NZ Gravity}%\noaffiliation

\date{\today}% It is always \today, today,

\begin{abstract}
 %This paper introduces a novel spectral density estimation method designed for analyzing very long multivariate time series, such as those encountered in data from gravitational wave detectors like the Einstein Telescope or LISA.
 %% AVI: I dont think its actually a novel method -- but a novel application... Hu + Prado introduced the novel method. We build upon the novel method + improve it with optimisation techniques + demonstrate with ET data.
Gravitational wave detectors like the Einstein Telescope and LISA generate long multivariate time series, which pose significant challenges in spectral density estimation due to a number of overlapping signals as well as the presence of correlated noise.
Addressing both issues is crucial for accurately interpreting the signals detected by these instruments.
This paper presents an application of a variational inference spectral density estimation method specifically tailored for dealing with correlated noise in the data. It is flexible in that it does not rely on any specific parametric form for the multivariate spectral density.
The method employs a blocked Whittle likelihood approximation for stationary time series and utilizes the Cholesky decomposition of the inverse spectral density matrix to ensure a positive definite estimator. 
A discounted regularized horseshoe prior is applied to the spline coefficients of each Cholesky factor, and the posterior distribution is computed using a stochastic gradient variational Bayes approach. 
This method is particularly effective in addressing correlated noise, a significant challenge in the analysis of multivariate data from co-located detectors. 
The method is demonstrated by analyzing \SI{2000}{\second} of simulated Einstein Telescope noise, which shows its ability to produce accurate spectral density estimates and quantify coherence between time series components. This makes it a powerful tool for analyzing correlated noise in gravitational wave data.
 % We demonstrate the method's efficacy using simulated Einstein Telescope noise, showing that the approach not only produces accurate spectral density estimates but also quantifies the coherence between time series components, offering a robust tool for analyzing correlated noise in gravitational wave data.
 %We illustrate this technique for ET noise characterisation and simultaneous extraction of a binary merger signal.  For computational feasibility we approximate the posterior distribution the variational Bayesian method, minimizing the KL divergence with the variational distribution. As the result, we can estimate the spectral density quite accurately and quickly. 
\end{abstract}

\maketitle

% Acronyms
\begin{acronym}
    \acro{GW}[GW]{gravitational-wave}
    \acro{GWs}[GWs]{gravitational-waves}
    \acro{ET}[ET]{Einstein Telescope}
    \acro{LISA}[LISA]{Laser Interferometer Space Antenna}
    \acro{CE}[CE]{Cosmic Explorer}
    \acro{PE}[PE]{parameter estimation}
    \acro{PSD}[PSD]{Power Spectral Density}
    \acro{DFT}[DFT]{Discrete Fourier Transform}
    \acro{SNR}[SNR]{Signal-to-Noise Ratio}
    \acro{BBH}[BBH]{binary black hole}
    \acro{VI}[VI]{variational inference}
    \acro{KL}[KL]{Kullback-Leibler}
    \acro{MCMC}[MCMC]{Markov chain Monte Carlo}
    \acro{LVK}[LVK]{LIGO--VIRGO--KAGRA}
    \acro{SGVB}[SGVB]{stochastic gradient variational Bayes}
\end{acronym}

\section{Introduction}

The next-generation \ac{GW} detectors, e.g. \ac{ET}~\cite{Punturo_2010}, \ac{CE}~\cite{CE_horizon_study}, and \ac{LISA}~\cite{LISA_science_case}, will usher in a transformative era of \ac{GW} astronomy. 
The increased sensitivity to \ac{GW} will lead to a much higher \ac{GW} detection rate, improving our understanding of astrophysical and cosmic phenomena, ranging from the dynamics of black hole mergers to the nature of the early universe~\cite{ET_science_case, Maggiore_2020_ET_science_case, Branchesi_2023_ET_science_case, CE_horizon_study, LISA_science_case}.

However, multivariate time series analysis of future GW detector data faces challenges in managing correlated multivariate noise~\cite{ET_design_report,LISA_red_book}.
\ac{ET} and \ac{LISA}, which consist of co-located interferometers, may experience correlated noise in their data streams~\cite{Janssens2023}. 
For \ac{ET}, this correlated noise can stem from seismic, Newtonian, and magnetic sources~\cite{Janssens_newtonian_seismic, PhysRevD.109.102002,Ball_lightning_strokes, Janssens_magnetic_noise}, while \ac{LISA}  may encounter noise from temperature variations and micro-thrusters~\cite{lisa_temp_noise,lisa_thrusters_noise}.

Ignoring correlated noise in analyses of multivariate \ac{GW} data can lead to biased parameter estimates and incorrect astrophysical conclusions (e.g., in analyzing the stochastic \ac{GW} background~\cite{Thrane_correlations_SGWB, Christensen_2019_SGWB, boileau2022figures} and transient signals from events like binary black hole mergers~\cite{Cireddu:2023:arXiv}). 
Recent studies have shown the need for methods that can handle correlated noise~\cite{Janssens2023}, but did not include a demonstration of Bayesian methods 
capable of effectively doing so. Other work, \cite{Cireddu:2023:arXiv} did demonstrate the capability to estimate the physical parameters of a compact binary coalescence event in a detector network with correlated noise. Their likelihood included the multivariate spectral density but they treated the spectral density as fixed with a focus on estimating the parameters of the binary chirp signal.  While this is an important step forward in the field, the authors of \cite{Cireddu:2023:arXiv} made a number of simplifying assumptions. First of all, they considered the correlated noise to be in phase. Secondly, they considered the detectors to be (almost) maximally correlated across the entire frequency spectrum. So far no Bayesian method has been suggested to estimate  multivariate spectral densities  of gravitational wave network data including correlated noise.

In this work, we demonstrate a method which is able to account for correlated noise as well as uncorrelated noise in a simultaneous and consistent approach. Additionally we make no assumptions about the phase of the different noise terms and model both real and imaginary cross-terms. While we demonstrate our method on a simplified noise example, the method forms the basis for future work with more complex noise scenarios, but this work shows the effectiveness of the chosen approach.
Additionally our proposed method allows for
 a flexible Bayesian approach for matrix-valued spectral density estimation, in a nonparametric way. Such a method is robust with respect to deviations from certain parametric shapes such as power laws in order to be able to capture all potential small- and large-scale noise characteristics.
Furthermore, the approach must exhibit computational efficiency when applied to long time series of \ac{GW} observations, as in the case of ET and LISA. 
\citet{MeierAlexander2020Bnao} and \citet{Liu2023} have developed non-parametric Bayesian approaches to multivariate spectral density estimation based on the Whittle likelihood and a Bernstein -- Hermitian positive semidefinite matrix-Gamma process prior on the spectral density matrix and used an adaptive MCMC algorithm to sample from the posterior. 
Although theoretically attractive with proven consistency properties and contraction rates, MCMC methods require significant computation time due to the complexity of sampling from high-dimensional posterior distributions, making them impractical for long time series of \ac{GW} observations, where the computational burden increases dramatically.

In this paper, a stochastic gradient variational Bayes (SGVB) approach, as developed by \citet{Hu2023}, is proposed to estimate the multivariate \ac{PSD} for correlated \ac{ET} noise. As a development within the broader framework of variational inference (VI), SGVB optimizes a surrogate posterior distribution by minimizing the \ac{KL} divergence from the true posterior distribution~\cite{Jordan1999,Wainwright2008,Blei2017}.  
This method transforms complex posterior sampling into an optimization problem, enhancing sampling efficiency and reducing computational demands~\cite{Blei2006,kingma2022}.
Recent advances in VI, such as normalizing flows, have demonstrated its effectiveness for pulsar timing-array datasets~\cite{Vallisneri2024}.

In this article, we demonstrate the usefulness and flexibility of a SGVB approach to estimate the spectral density matrix of the correlated instrumental noise from a  network of \ac{GW} detectors.
In particular, the approach is illustrated by using simulated ET noise and is also readily applicable to estimate  \ac{LISA} instrumental noise or correlated \ac{LVK} noise induced by magnetic noise such as from the Schumann resonances~\cite{Thrane_correlations_SGWB}.
To ease computational efforts of handling large time series, a blocked Whittle likelihood approximation is introduced and utilized instead of the traditional Whittle likelihood. 
Advice is provided on choosing tuning parameters such as the learning rate and number of basis functions in the SGVB approach. 
Finally, as a by-product of this method, the squared coherence is computed  as a function of frequency, quantifying the degree of inter-temporal correlation between components of the multivariate time series. 

The article is structured as follows.
Section~\ref{sec:method} introduces the \ac{SGVB} method and defines the blocked Whittle likelihood.
The efficiency and accuracy of the SGVB approach are tested using simulations in Section~\ref{sec:simulation}. 
Section~\ref{sec:application} presents results of the method applied to simulated \ac{ET} datasets consisting of varying levels of correlated noise. 
Lastly, the article concludes with a summary of findings in Section~\ref{sec:Discussion}.

\section{Method}
\label{sec:method}

\subsection{Likelihood}

Assume we are given a time series of gravitational wave observations of length $n$ from $p$ channels. To set notation,
let $\Z=(\Z_1,\ldots,\Z_n)^ \intercal\in  \mathbb{R}^{n\times p}$ be a $p$-dimensional stationary, mean-zero time series, sampled at time intervals $\Delta_t=1/(2f_{Ny})$, where $f_{Ny}$ is the Nyquist frequency, for a total observation time of $T$ with a total of $n=T/\Delta_t$ sampled values for each of the $p$ dimensions. The frequency resolution, \(\Delta_f\), is given by
\begin{align}
\Delta_f = \frac{1}{n \Delta_t} = \frac{1}{T}\, .
\end{align}
Let $\d_k$ be the \ac{DFT} of $\Z$ given by
\begin{align}
\d(f_k) = \sqrt{\frac{\Delta_t}{n}}\sum_{t=1}^{n} \Z_t\exp \left(-2\pi i \frac{k}{n} t \right)\, ,
\end{align}
where $f_k= k \Delta_f= k/({n\Delta_t})=k /{T}$ for $k=1,\ldots, N$, where $N = n/2$ if n is even, $N = (n-1)/2$ if n is odd. In this analysis, we use the two-sided Fourier transform, which includes both positive and negative frequencies, allowing for a full representation of the frequency spectrum. However, for the purposes of this study, we only retain the positive frequency components.

Due to the normalizing and decorrelating properties of the Fourier transform, the discrete Fourier coefficients $\d(f_k)$ %(multiplied by $1 / \sqrt{n}$)
are asymptotically independent and have a complex Gaussian distribution with covariance matrix given by the spectral density matrix $\S(f_k)$,  the Fourier transform of the autocovariance function. This asymptotic Gaussian distribution is the basis of the multivariate Whittle likelihood approximation in the frequency domain, expressed as
\begin{align}\label{eq:Whittle likelihood}
 \mathcal{L}(\d|\S) &\propto  \prod_{k=1}^{N} \det(\S(f_k))^{-1} \times \nonumber \\
 & \exp\left(-\d(f_k)^* \S(f_k)^{-1} \d(f_k)\right)
\end{align}
where $\d(f_k)^*$ is the conjugate transpose of the $\d(f_k)$ and $\S(f_k)$ is a $p \times p$ Hermitian positive semidefinite spectral density matrix at each $f_k$. Therefore, the unknown quantity here is $\S$, a matrix-valued function at each frequency with the additional restriction that its value at each frequency is Hermitian positive semidefinite. Note that when estimating the spectral density matrix, it is important that the estimate is again positive semidefinite at each $f_k$ so that the quadratic form in the exponent of the Whittle likelihood remains positive and thus defines a valid likelihood.

Given $\d$, one method to estimate the multivariate spectral density matrix $\S$ is Bayesian inference. In particular, having a flexible Bayesian method instead of simply using a frequentist Welch estimate is important when the ultimate task is to simultaneously estimate the parameters of a GW signal while properly taking the uncertainties of the instrumental noise estimation into account. The Bayes' theorem is given by
\begin{align}
    p(\S|\d) =& \frac{\mathcal{L}(\d|\S)\pi(\S)}{\mathcal{Z}(\d)} \nonumber \\
    \propto& \mathcal{L}(\d|\S)\pi(\S)\, ,
\end{align}
where $\pi(\S)$ is the prior distribution and $p(\S|\d)$ is the posterior density of unknown $\S$ given $\d$, 
and $\mathcal{Z}(\d)$ is the Bayesian evidence (see ~\citet{thrane_talbot_bayesian_primer} and ~\citet{Christensen_PE_for_GW} for discussions on \ac{GW} Bayesian inference).

In GW parameter estimation with large datasets, such as those from ET and LISA, the computational burden of directly applying the Whittle likelihood can be significant. To mitigate this challenge, a `blocked' Whittle likelihood is adopted~\cite{vu2024}. 
The time series is divided into $N_{b}$ equal-sized blocks $\Z=(\Z^{(1)},\ldots,\Z^{(N_{b})})^\intercal $ with each block a $p$-dimensional time series of each of length $n/N_{b}$. The discrete Fourier transform of each block is denoted by 
$\d^{(i)}$, $i=1,\ldots,N_{b}$. The stationarity assumption implies that the statistical properties of each block, in particular, their spectral densities are the same.
Assuming independence among different blocks, the likelihood then becomes the product of the individual blocked Whittle likelihoods,
\begin{equation}\label{eq:block_lnl}
    \mathcal{L}_b(\d|\S) = \prod^{N_b}_{i=1} \mathcal{L}(\d^{(i)}|\S) \ .
\end{equation}
Note that as the length of the blocked data is less than the original dataset, the frequency resolution of the spectral density will become coarser as the number of blocks $N_b$ increases. 
In practice, the number of blocks should be chosen to achieve a required frequency resolution while also remaining computationally tractable. 

\subsection{Parametrization of $\S$}
\label{subsec: parameters}
The prior defined by \citet{RosenOri2007Aeom,Hu2023} models the components of the Cholesky factorization of the inverse spectral density matrix  via smoothing splines (for details, refer to \citet{Hu2023}).
In their approach, the inverse of $\S(f_k)$ can be represented as $\S(f_k)^{-1} = \mathbf{T}_k^* \mathbf{D}_k^{-1} \mathbf{T}_k$, where $\mathbf{D}_k$ is a diagonal matrix with diagonal elements $\delta_{1k}^2, \delta_{2k}^2, ..., \delta_{pk}^2$ and
\begin{align}
\mathbf{T}_k = \begin{pmatrix}
1 & 0 & 0 & \cdots & 0 \\
-\theta_{21}^{(k)} & 1 & 0 & \cdots & 0 \\
-\theta_{31}^{(k)} & -\theta_{32}^{(k)} & 1 & \ddots & \vdots \\
\vdots & \vdots & \ddots & \ddots & 0 \\
-\theta_{p1}^{(k)} & -\theta_{p2}^{(k)} & \cdots & -\theta_{p,p-1}^{(k)} & 1
\end{pmatrix}
\end{align}
is a $p \times p$ complex unit lower triangular matrix. Thus the Whittle likelihood can be rewritten as the product of its $p$ components depending on $\btheta_j,\bdelta_j$, for $j=1,\ldots,p$,
\begin{align}
 \mathcal{L}(\d|\S)
 \propto \prod_{j=1}^{p} \mathcal{L}_j(\d_j|{\btheta_j,\bdelta_j})
 \label{eq:likelihood}
\end{align}
where
\begin{align}
\mathcal{L}_j(&\d_j|\btheta_j,\bdelta_j) \propto \nonumber \\
&  \prod_{k=1}^{N} \delta_{jk}^{-2} \exp \left(\frac{-\left|d_j(f_k)-\sum_{l=1}^{j-1}\theta_{jl}^{(k)}d_l(f_k) \right|^2}{\delta_{jk}^2} \right)
\label{eq:likelihoodj}
\end{align}
the parameters $\btheta_j,\bdelta_j$ represent the set of $\theta_k$ and $\delta_k$ for the $j$th component of the multivariate time series, $\theta_{jl}^{(k)}$ represents the corresponding element in the matrix $T_k$ for any $j>l$, and $\theta_{jl}^{(k)} = 1$ when $j=l$, $d_j(f_k)$ denotes the Fourier coefficient of the $j$th component, and $\d_j=(d_j(f_1),\ldots,d_j(f_N))^\intercal$.
Then $\log(\delta_{jk}^2)$ and the real and imaginary parts of $\theta_{jl}^{(k)}$ are modeled by Demmler-Reinsch basis functions in terms of $M$ truncated smoothing splines, given by

\begin{align}
\Re(\theta_{jl}^{(k)}) &= \alpha_{jl,0} + \alpha_{jl,1}f_k + \sum_{s=1}^{M-1}\psi(f_k)\alpha_{jl,s+1}, \label{eq:real_theta} \\
\Im(\theta_{jl}^{(k)}) &= \beta_{jl,0} + \beta_{jl,1}f_k + \sum_{s=1}^{M-1}\psi(f_k)\beta_{jl,s+1}, \label{eq:imag_theta} \\
\log \delta_{jk}^2 &= \gamma_{j,0} + \gamma_{j,1}f_k + \sum_{s=1}^{M-1}\psi(f_k)\gamma_{j,s+1},  \label{eq:log_delta}
\end{align}
where $\psi(x) = \sqrt{2} \cos(s\pi x)$, with $s$ being an integer that ranges from 1 to $M-1$, represents the spline basis function. 
The flexibility of the model can be adjusted by choosing the number of basis splines $M$.
Thus, a flexible model for the spectral densities $\S=\S(\bnu)$ depending on a parameter vector $\bnu$  that comprises all $\alpha, \beta$ and $\gamma$ parameters is constructed. 
Following \citet{Hu2023}, discounted regularized horseshoe prior\cite{PiironenJuho2017Siar} are used for the $\alpha, \beta$ and $\gamma$ spline coefficients.
This prior is adept at handling varying degrees of smoothness in the individual components of the spectral density matrix while avoiding overfitting (for details, refer to \cite{Hu2023,PiironenJuho2017Siar}).
Thus, in line with Equation~\ref{eq:likelihood}, the subvector of the parameter vector $\bnu$ that contains all parameters for component $j$ is denoted as $\bnu_j$.
This allows us to decompose the posterior into the product of $p$ posteriors for each individual component, given by
\begin{align}
p(\bnu|\d )= \prod_{j=1}^p p_j(\bnu_j|\d_j).
\end{align}
This prescription allows the application of the SGVB approach to each $p_j(\bnu_j|\d_j)$ in parallel. 
The approach can be futher parallelised by computing each of the likelihood `blocks' from Equation~\ref{eq:block_lnl} independently.

\subsection{Stochastic Gradient Variational Bayes}
\label{subsec:sgvb_details}

This section provides a brief review of SGVB, along with discussions on how to tune the learning rate and the number of basis splines.\smallskip

\paragraph{Variational Inference Review:}
The fundamental concept of \ac{VI} is to approximate the posterior distribution $p_j(\bnu_j|\d_j)$ by using a surrogate distribution from a family of variational distributions ${\cal Q}_j=\{ q_{\phi_j}(\bnu_j); \phi_j \in \Phi_j \}$, which depends on a parameter vector $\phi_j$ within a parameter space $\Phi_j$.
Typically, the variational family is selected for its simplicity and computational tractability. 
A product of normal distributions with mean $\mu_{ji}$ and variance $\sigma^2_{ji}$ is utilized for each element of the parameter $\bnu_j$.
The goal of the variational approach is to identify the parameters $\phi_j=(\mu_{ji},\sigma^2_{ji})$, for $i=1,\ldots,\mbox{dim}(\bnu_j)$, that minimize the reverse Kullback-Leibler (KL) divergence between the variational family and the true posterior distribution, denoted as $d_{KL}(q_{\phi_j}||p_j)$), i.e.,
\begin{align}\label{eq:phi_min}
  \phi_j^* &= \operatorname*{argmin_{\phi_j}} d_{KL}(q_{\phi_j}||p_j) \nonumber \\
  &= \operatorname*{argmin_{\phi_j}} \int \log\frac{q_{\phi_j}(\bnu_j)}{p_j(\bnu_j|\d_j)}q_{\phi_j}(\bnu_j) \, d\bnu_j \, .
\end{align}

The optimization algorithm employed in this work uses a SGVB approach, as described by \citet{kingma2022,Xu2019,Domke2019}. \smallskip

\paragraph{Choice of the Number of Basis Functions:}

The variation of parameters ($\log \delta^2_{jk},\Re(\theta^{(k)}_{j\cdot}),\Im(\theta^{(k)}_{j\cdot})$) across frequencies $f_k$ is modeled using $M$ truncated smoothing splines. 
Choosing an appropriate $M$ is crucial: a low $M$ may cause underfitting, while a high $M$ can lead to overfitting and increased computational complexity due to the higher dimensionality of $\phi_j$. To address overfitting concerns, a horseshoe prior on the coefficients acts as a regularization mechanism, effectively reducing the risk of overfitting~\citep{10.1214/17-EJS1337SI}.

The likelihood function (Equation~\ref{eq:likelihoodj}) measures the consistency of the data with the spectral density, parameterized by $(\log \delta^2_{jk}, \Re(\theta^{(k)}_{j\cdot}), \Im(\theta^{(k)}_{j\cdot}))$. 
We use the likelihood in the maximum likelihood estimate (MLE) to choose $M$. 
The likelihood at MLE tends to increase with $M$, reflecting an improvement in model fit as the model complexity increases. 
When $M$ is sufficiently large, the likelihood tends to stabilize or increase steadily. 
We chose $M$ to prevent underfitting by conservatively selecting a slightly larger $M$ than the point where the likelihood stabilizes.

\smallskip

\paragraph{Optimization of the Learning Rate:}\label{subsec:learningrate}

The KL minimizer $\phi^*_j$ (Equation~\ref{eq:phi_min}) is equivalent to the maximizer of the evidence lower bound (ELBO) between $q_{\phi_j}$ and $p_j(\bnu_j|\d_j)$ i.e.,  
\begin{align}\label{eq:elbo}
\text{EL}&\text{BO}(q_{\phi_j},\ p_j(\bnu_j|\d_j))  \nonumber \\
&=\mathbb{E}_{\bnu_j\sim q_{\phi_j}(\bnu_j)}[\log p_j(\bnu_j|\d_j)-\log q_{\phi_j}(\bnu_j)] \,.    
\end{align}
The SGVB approach is used to estimate $\phi^*_j$, $j=1,...,p$ and it consists of the next two steps:
\begin{enumerate}
    \item Maximize $\log p(\boldsymbol{\nu}_j|\mathbf{d}_j)$ with respect to $\boldsymbol{\nu}_j$.
    \item Maximize the ELBO (Equation~\ref{eq:elbo}) with respect to $\phi_j$ after substituting the maximized $\log p(\boldsymbol{\nu}_j|\mathbf{d}_j)$ from the first step.
 \end{enumerate}

The performance of each maximizer depends on the learning rate; if the learning rate is too small or too large, the optimization algorithm may get stuck at a local maximum or take too long to find the global maximum. 
Thus, choosing the optimal learning rate is crucial. Recently \cite{Welandawe2024} proposed automated techniques for the blackbox VB in which the learning rate is adaptively decreased. The reliability of the stochastic optimization methods for VB due to substantial hand-tuning parameters to apply effectively were discussed in literature \cite{Agrawal2020,Welandawe2024}.

While \citet{Hu2023} selected a specific but arbitrary value for the learning rate, this study proposes a method to select the optimal learning rate and automate the selection procedure.

Let $\tau_1$ and $\tau_2$ be the learning rates associated with the first and second steps, respectively. 
As $\tau_2$ has a broad range of plausible values, it is unnecessary to deviate from the default settings proposed by \citet{Hu2023}. 
An appropriate $\tau_1$ is obtained by maximizing the ELBO 
\begin{equation}
\tau_1^* = \operatorname*{argmax_{\tau_1}} \prod^p_{j=1}\text{ELBO}(q_{\phi_j}, p_j(\boldsymbol{\nu}_j|\mathbf{d}_j)) \,.    
\end{equation}
This optimization is carried out over a continuous parameter space using the Python package \hyperopt~\cite{Bergstra2013} and the tree-structured Parzen Estimator (TPE) algorithm~\cite{Bergstra2011}. 
From our preliminary study, the optimal learning rates obtained by fixing $\tau_2$ coincide with those obtained when optimizing $\tau_1$ and $\tau_2$ jointly.
The maximum number of iterations for the optimization procedure is set to \num{10 000}.

\subsection{Squared Coherence}

To quantify the frequency-dependent relationship between the channels of the multivariate ET time series, we employ the squared coherence, widely used in many fields~\cite[e.g.,][]{Sakkalis2011,wiley1969}. 
Squared coherence, $C_{xy}(f_k)$, is a normalized measure of association between two time series at frequency $f_k$, ranging from 0 (no correlation) to 1 (perfect correlation). 
For two components of a multivariate time series, it is defined as:

\begin{align}\label{squared coh}
C_{xy}(f_k) = \frac{|\S_{xy}(f_k)|^2}{\S_{xx}(f_k)\S_{yy}(f_k)}
\end{align}
where $\S_{xy}(f_k)$ is the cross-spectral density between components $x$ and $y$, and $\S_{xx}(f_k)$, $\S_{yy}(f_k)$ are the spectral densities of components $x$ and $y$, respectively, at frequency $f_k$ $(x,y = 1,2,...,p; x\neq y)$, for $x,y = 1,2,...,p,$ with $x\neq y$.

\section{Simulation study}
\label{sec:simulation}

In order to test the efficiency and accuracy of SGVB, 500 independent realisations of a bivariate time series are generated from a vector autoregressive model of order 2 (VAR(2)) and a vector moving-average model of order 1 (VMA(1)), using three  different sample sizes $n=256;512;1024$ (refer to \citet[Section~4.2,][]{Liu2023} for the definitions of the VAR(2) and VMA(1) models).
The spectral densities are estimated using both the \ac{SGVB} method and \citet{Liu2023}'s vectorized nonparametrically corrected method (VNPC), an MCMC-based approach that samples from the exact posterior distribution without a variational approximation\footnote{For the VNPC analyses, we use the software and settings provided by \citet{Liu2023}, specific for the VAR(2) and VMA(1) models.}.

\begin{figure}
  \centering
  \includegraphics[width=0.85\columnwidth]{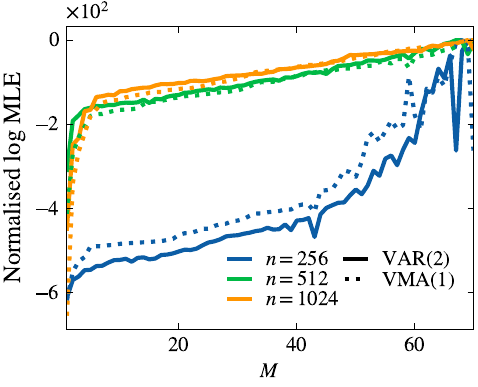}
  \captionof{figure}{
  The relationship between the number of basis functions ($M$) and the log-normalized likelihood at the MLE for VAR(2) and VMA(1) models (solid and dotted lines respectively) with different data lengths $n=256$ (blue), $n=512$ (green), and  $n=1024$ (orange). A fixed learning rate of 0.002 was used across all cases.
}
  \label{fig:sim_basis}
\end{figure}

In order to avoid underfitting in SGVB, an appropriate number of basis functions is determined for each dataset using the method discussed in Section~\ref{subsec:sgvb_details}. 
Figure~\ref{fig:sim_basis} shows the log MLE (normalized for comparison between datasets) plotted against the number of basis functions.
$M=30$ is set as the log MLE enters a steady ascent phase and before any perturbations appear, thereby avoiding both underfitting across the three different dataset sizes. 
When $n=256$, a rapid increase of the log-likelihood is observed as $M>50$, indicating overfitting. As shown in equations \eqref{eq:real_theta}, \eqref{eq:imag_theta}, and \eqref{eq:log_delta}, the number of parameters is $4(M+2)$ for a bivariate time series, which exceeds the number of data points.

\begin{figure}
  \centering
  \includegraphics[width=0.9\columnwidth]{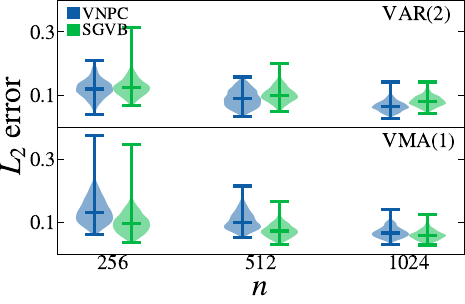}
  \captionof{figure}{Violin plots of VNPC and SGVB $L_2$ errors from 500 realizations for VAR(2) and VMA(1) models, for different data lengths $n$. The SGVB method uses optimized values for $M$ and $\tau_1$.}
  \label{fig:sim_error_violins}
\end{figure}

\begin{figure}
  \centering
  \includegraphics[width=0.9\columnwidth]{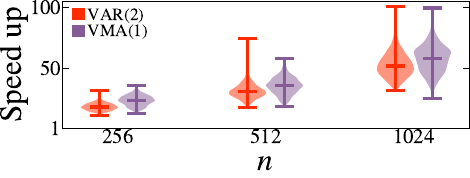}
  \captionof{figure}{Violin plots of the speed up gained by using SGVB compared to VNPC from 500 realizations for VAR(2) and VMA(1) models, for different data lengths $n$. The SGVB method uses optimized values for $M$ and $\tau_1$.}
  \label{fig:sim_speed_violins}
\end{figure}

Given an estimated spectral density matrix $\hat{\S}$ and the true spectral density matrix $\S_0$, the accuracy of PSD estimates can be assessed using the $L_2$ error. The $L_2$ error provides a quantitative measure of the discrepancy between  $\hat{\S}$ and $\S_0$, encapsulating the overall difference between the estimated and true spectral density matrices across the frequency range. 
A smaller $L_2$ error indicates that $\hat{\S}$ closely approximates $\S_0$, while a larger $L_2$ error reflects greater deviation.
The $L_2$ error can be approximated as
\begin{equation}
 ||\hat{\S} - \S_0||_{L_2}  \approx \left(\frac{1}{N} \sum_{k=1}^{N}||\hat{\S}_0(f_k)-\S_0(f_k)||^2 \right)^{1/2}\, ,
\end{equation}
where $\Vert \cdot \Vert$ denotes the Frobenius norm, which  for a complex $p\times p$ matrix $\A$ is defined by
\begin{equation}
\Vert \A\Vert= \sqrt{\sum_{i,j} |a_{ij}|^2}\, .
\end{equation}

Figure~\ref{fig:sim_error_violins} displays the $L_2$ error distributions from both the SGVB and VNPC methods across the different sample sizes. 
The $L_2$ errors are calculated by using the estimated spectral densities and the true spectral densities of the VAR(2) simulations (top panel) and the VMA(1) simulations (bottom panel), respectively.

Figure~\ref{fig:sim_speed_violins} displays the speed-up factor of the SGVB compared to the VNPC method. 
Additionally, Table~\ref{table:simstudy} presents the simulation study's median $L_2$ errors, pointwise coverage, and median width of pointwise 90\% credible regions, as well as the median computation time (in seconds)\footnote{Pointwise coverage is defined as the proportion of times the true spectral density value at each frequency is captured within the confidence interval across different realizations of the data.}.

\begingroup
\begin{table*}[!htbp]
\centering
\renewcommand{\arraystretch}{1.5}
\setlength{\tabcolsep}{3pt}
\begin{NiceTabular}{l l *{3}{|cc}}[colortbl-like, cell-space-limits=1pt]
\CodeBefore
  \rowcolors{5}{gray!10}{white}
  \columncolor{white}{1-2}
\Body
\Block{2-8}{} \\
  &   & \Block{1-2}{$n=256$} & & \Block{1-2}{\bfseries $n = 512$} & & \Block{1-2}{\bfseries $n = 1024$} & \\
& & VNPC & SGVB & VNPC & SGVB & VNPC & SGVB\\
\hline
\Block{7-1}{\bfseries VAR(2)}
& $L_2$ error & 0.12 ± 0.02 & 0.12 ± 0.02 & 0.10 ± 0.02 & 0.09 ± 0.02 & 0.08 ± 0.01 & 0.06 ± 0.01 \\
& Pointwise coverage & 0.87 ± 0.06 & 0.62 ± 0.11 & 0.85 ± 0.06 & 0.66 ± 0.10 & 0.84 ± 0.06 & 0.67 ± 0.07 \\
& $\S_{11}$ CI Width & 0.09 ± 0.01 & 0.06 ± 0.01 & 0.06 ± 0.01 & 0.04 ± 0.01 & 0.04 ± 0.00 & 0.03 ± 0.00 \\
& $\Re \S_{12}$ CI Width & 0.09 ± 0.01 & 0.06 ± 0.01 & 0.07 ± 0.01 & 0.04 ± 0.01 & 0.05 ± 0.00 & 0.03 ± 0.00 \\
& $\Im \S_{12}$ CI Width & 0.07 ± 0.01 & 0.04 ± 0.01 & 0.06 ± 0.01 & 0.04 ± 0.01 & 0.04 ± 0.00 & 0.03 ± 0.00 \\
& $\S_{22}$ CI Width & 0.13 ± 0.01 & 0.07 ± 0.01 & 0.09 ± 0.01 & 0.05 ± 0.01 & 0.07 ± 0.01 & 0.04 ± 0.00 \\
& Time [s] & 5.4K ± 0.6K & 280 ± 30 & 9.4K ± 1.3K & 300 ± 30 & 18.4K ± 2.3K & 350 ± 30 \\
\hline 
\Block{7-1}{\bfseries VMA(1)}
& $L_2$ error & 0.10 ± 0.03 & 0.13 ± 0.03 & 0.07 ± 0.02 & 0.10 ± 0.02 & 0.06 ± 0.01 & 0.07 ± 0.01 \\
& Pointwise coverage & 0.92 ± 0.07 & 0.64 ± 0.10 & 0.92 ± 0.08 & 0.63 ± 0.09 & 0.87 ± 0.10 & 0.67 ± 0.10 \\
& $\S_{11}$ CI Width & 0.12 ± 0.02 & 0.09 ± 0.02 & 0.08 ± 0.01 & 0.07 ± 0.01 & 0.06 ± 0.01 & 0.05 ± 0.01 \\
& $\Re \S_{12}$ CI Width & 0.08 ± 0.01 & 0.06 ± 0.01 & 0.05 ± 0.01 & 0.04 ± 0.01 & 0.04 ± 0.00 & 0.03 ± 0.00 \\
& $\Im \S_{12}$ CI Width & 0.07 ± 0.01 & 0.04 ± 0.01 & 0.06 ± 0.01 & 0.03 ± 0.00 & 0.03 ± 0.00 & 0.03 ± 0.00 \\
& $\S_{22}$ CI Width & 0.17 ± 0.03 & 0.18 ± 0.04 & 0.12 ± 0.02 & 0.13 ± 0.02 & 0.08 ± 0.01 & 0.10 ± 0.01 \\
& Time [s] & 4.2K ± 0.3K & 180 ± 30 & 6.9K ± 0.5K & 190 ± 30 & 12.6K ± 1.0K & 220 ± 40 \\
\end{NiceTabular}
\caption{Comparison of median $L_2$ errors, empirical pointwise coverage, median width of pointwise 90\% credible intervals, and median computation time (in seconds) with their respective MAD (median absolute deviation) for 500 simulations using VNPC and SGVB methods across different sample sizes ($n=256$, $512$, and $1024$) for both VAR(2) and VMA(1) models.}
\label{table:simstudy}
\end{table*}
\endgroup

Figure~\ref{fig:sim_error_violins} reveals that as the sample size increases, both methods achieve lower median $L_2$ errors, and the median $L_2$ errors are similar between the two methods. 
Additionally, Figure~\ref{fig:sim_speed_violins} shows the median speed-up factor for SGVB becomes increasingly pronounced, ultimately surpassing VNPC by more than 50 times, as $n$ is increased. 
Notably, when $n=1024$, SGVB's accuracy is only marginally lower than VNPC's, yet it reduces median computation time by a factor of approximately 50. 
The simulation study provides empirical evidence for the consistency of the SGVB approach, demonstrating that as sample size grows, the posterior distributions contracts around the true spectral density, following the same trend as the VNPC approach. However, it does so at a fraction of the computation time.

Finally, Table~\ref{table:simstudy} illustrates that, even though VNPC method consistently outperforms SGVB across all sample sizes in terms of pointwise coverage, typically achieving 83-90\% coverage compared to SGVB's 61-66\%, SGVB generally produces narrower credible regions.
%, suggesting a trade-off between coverage and precision in the estimates.
%
The limited coverage and narrow credible intervals observed from the SGVB results can be attributed to the mean field approximation~\cite{Blei2017} which assumes independent components ignoring correlations between parameters, and using the KL divergence as distance measure. 
The combined effects tend to underestimate the marginal variances of the target distribution~\cite{Blei2006,Wang2005}. 
To address these limitations, researchers have explored more flexible surrogate distributions and a broader range of divergence measures in variational inference\cite{Kingma2013, Blei2017, Rezende2015,Yang2020,Higgins2017}. 

Additionally, studies have investigated model misspecification and developed assessment methods for approximation inference~\cite{Wang2019, yao2018, lee2019, yu2021}.
While more sophisticated frameworks yield improved posterior approximations, they often demand greater computational resources. 
In this work, we prioritize computational efficiency while maintaining an acceptable level of accuracy.

\section{Application to ET noise}
\label{sec:application}
\subsection{Data Generation}
\label{sec:data_gen}

Three independent realizations of Gaussian noise were synthesized (one for each of the XYZ channels), each spanning \SI{2000}{\second}, spectrally shaped to match the design sensitivity of the ET xylophone configuration~\cite{Hild_2009,Hild:2010id} 
\footnote{Note that the sensitivity curve used here is the previously called `ET-D' sensitivity curve. We choose this curve rather than the updated version presented by ~\citet{Branchesi:2023mws}, to allow a direct comparison with previous work \cite{Janssens2023}. 
The small difference between PSD curves will have no impact on the key conclusions of the applicability of the work presented in this paper.
}.
The noise was sampled at a frequency of \SI{2048}{Hz}, resulting in a multivariate time series comprising \SI{4096}{\kilo} data points per channel. 
In the rest of the paper we refer to this colored Gaussian noise as \textit{ET noise}.

Following \citet{Janssens2023}, non-identical correlated noise in the X, Y, and Z channels is simulated by injecting colored Gaussian noise characterized by frequency-domain Gaussian peaks. 
The power spectral density is given by 
\begin{equation}
    \S_n^{GP}(f, \mu, A) = \left( \frac{A}{\sqrt{2\pi}} \exp\left\{ -\frac{1}{2}  (f-\mu)^2 \right\} \right)^2\, ,
\end{equation}
where $A$ represents the amplitude, and $\mu$ the frequency peak location.

Gaussian noise is injected at specific frequencies in each channel as follows
{\small 
\begin{align*}
    \textbf{X}&: \textstyle{\S_n^{GP}(\mu\eqq\SI{10}{Hz}, A\eqq4\ex{24}), \S_n^{GP}(\mu\eqq\SI{50}{Hz}, A\eqq2\ex{24})}\, ,\\
    \textbf{Y}&: \S_n^{GP}(\mu\eqq\SI{10}{Hz}, A\eqq4\ex{24}), \S_n^{GP}(\mu\eqq\SI{90}{Hz}, A\eqq1.5\ex{24})\, ,\\
    \textbf{Z}&: \S_n^{GP}(\mu\eqq\SI{50}{Hz}, A\eqq2\ex{24}),  \S_n^{GP}(\mu\eqq\SI{90}{Hz}, A\eqq1.5\ex{24}).
\end{align*}
}
To introduce correlated noise between channel pairs, we utilize identical Gaussian peaks at matching frequencies across the paired channels. 
In contrast, for uncorrelated noise scenarios, independent Gaussian peaks are simulated for each channel, ensuring that there is no cross-channel correlation.

As stated in \citet{Janssens2023}, this dataset is inherently simplified due to the distinct nature of the correlated noise terms, which are highly differentiable.
Nevertheless, it serves as a valuable initial demonstration of our approach.
Future work aims to investigate more realistic and complex noise scenarios, such as the presence of correlated magnetic \citet{Janssens_magnetic_noise,Ball_lightning_strokes} and/or Newtonian noise \cite{Janssens_newtonian_seismic,PhysRevD.109.102002} as well as (multiple) GW signals. However, this is considered beyond the scope of this work where we want to lay the foundation for the estimation framework.

\subsection{Data Analysis}

\begin{figure}[!t]
\centering
  \includegraphics[width=0.85\columnwidth]{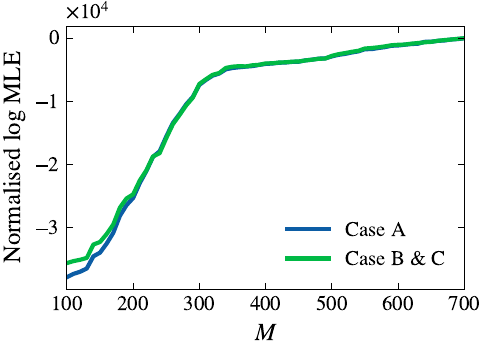}
  \caption{Relationship between the number of basis functions ($M$) and the normalized log maximum likelihood estimate (MLE) for different scenarios of Einstein Telescope noise analysis. Case A (blue line) represents correlated noise, while Cases B and C (green line) represent uncorrelated noise scenarios.
  }
  \label{et_corr_basis_funs_vs_mle}
\end{figure}

Three analyses are performed to estimate the power spectral density (PSD) and the corresponding squared coherences under the following different noise conditions, 
\begin{itemize}
\setlength{\itemindent}{-15pt}
    \item[] \textbf{Case A:}  ET noise with correlated Gaussian peaks,
    \item[] \textbf{Case B:} ET noise with uncorrelated Gaussian peaks,
    \item[] \textbf{Case C:} ET noise with uncorrelated Gaussian peaks, assuming independence between channels.
\end{itemize}

The data is divided into 125 equal blocks, each consisting of \num{16 384} frequency points and a maximum frequency of \SI{1024}{Hz} for the blocked Whittle likelihood estimation (Equation~\ref{eq:block_lnl}). 
Each Whittle likelihood block uses identical basis function expressions defined in Section~\ref{subsec: parameters} for the spectral density matrix. 

\begin{figure*}
\centering
\includegraphics[width=\textwidth]{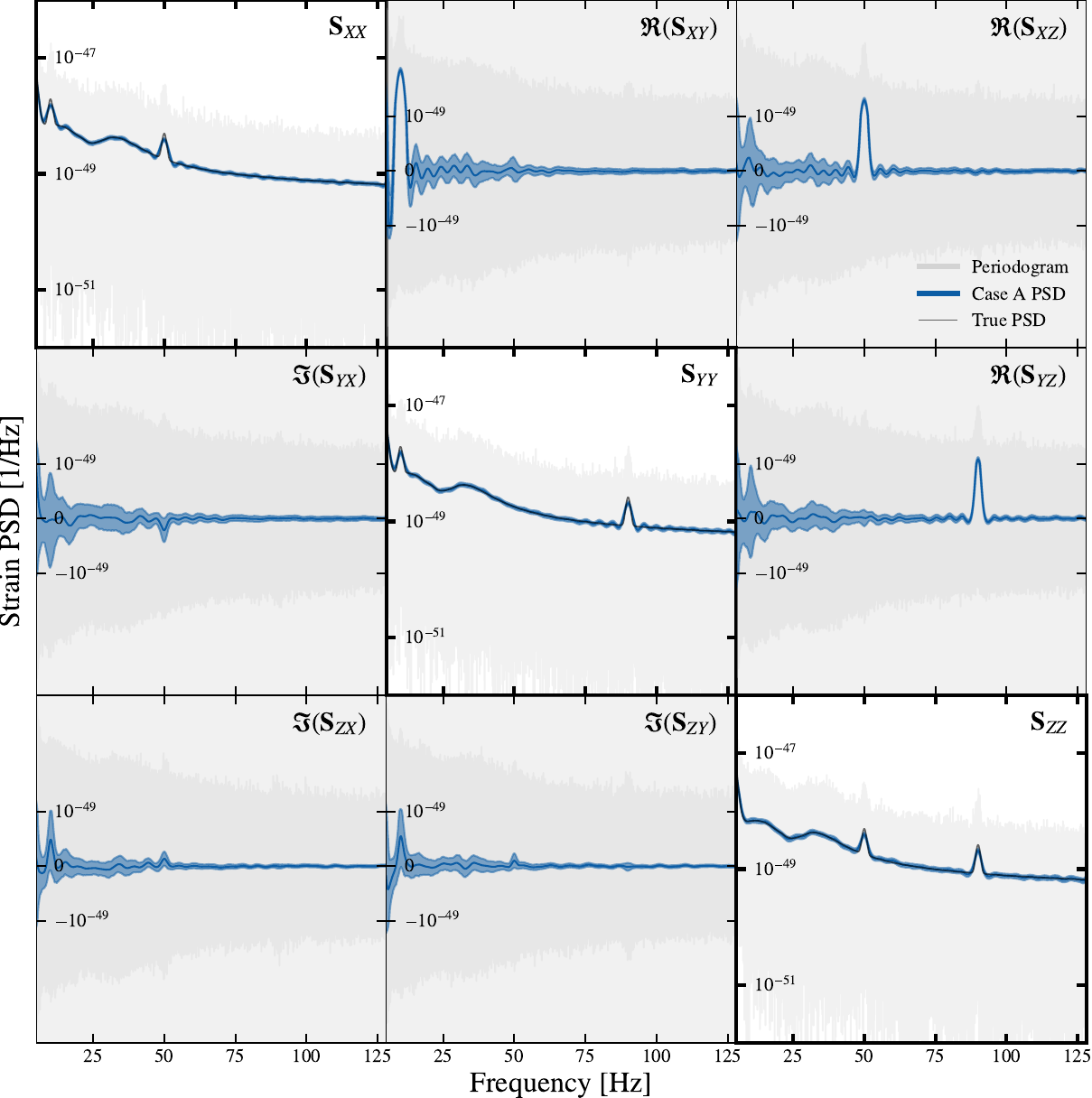}
\caption{Case A PSD. The plot shows the periodogram (gray), the PSD estimates (colored, with 90\% credible interval shaded) and the true PSD (black curve in the diagonals). The upper triangle subplots show the real parts of the off-diagonal elements of the spectral density matrix, while the lower triangle subplots show the imaginary parts of the off-diagonal elements.}
\label{fig:caseA_psd}
\end{figure*}

\begin{figure*}
\centering
\includegraphics[width=\textwidth]{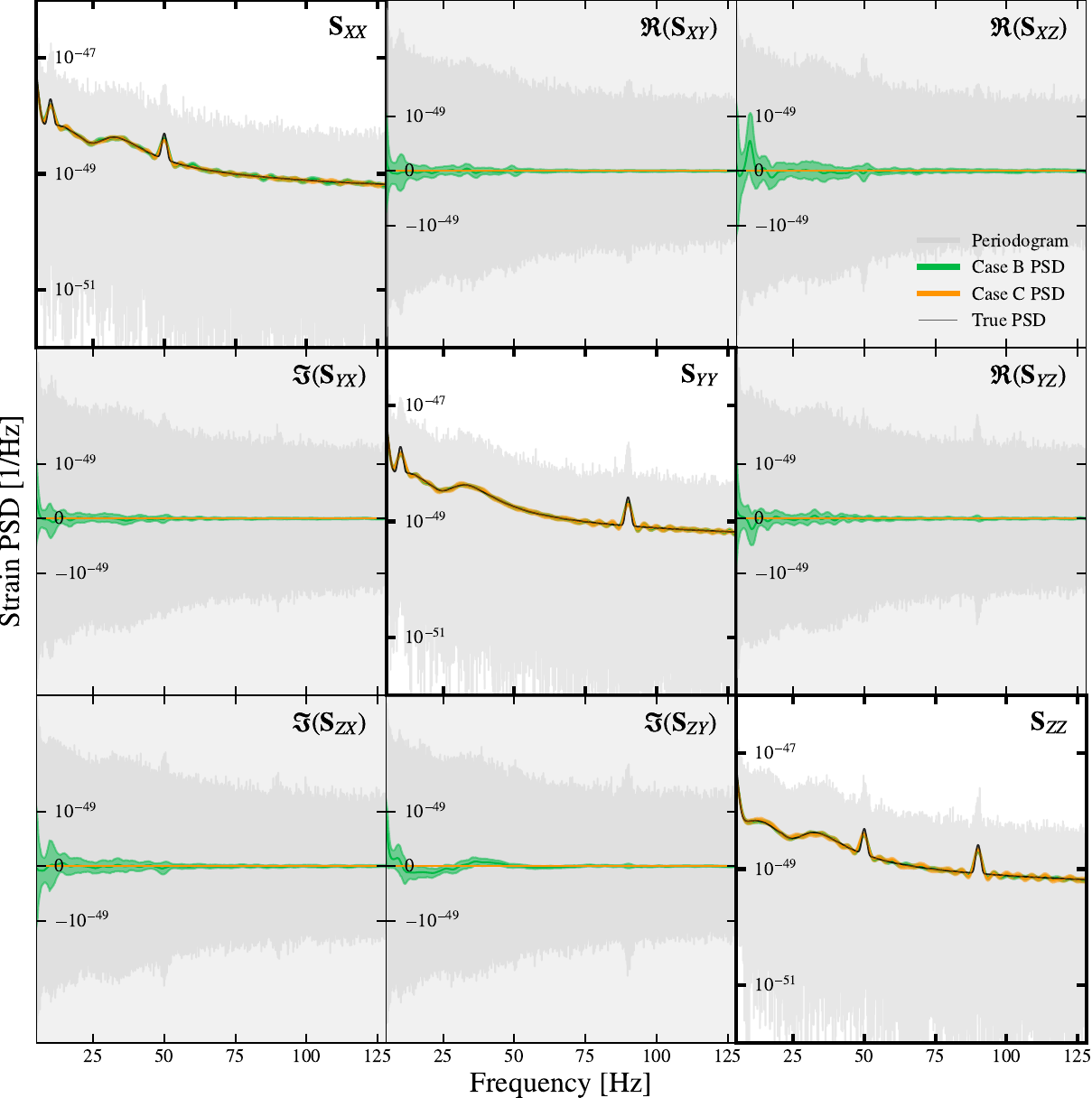}
\caption{Case B and C PSDs. The plot shows the periodogram (gray), the PSD estimates (colored, with 90\% credible interval shaded) and the true PSD (black curve in the diagonals). The upper triangle subplots show the real parts of the off-diagonal elements of the spectral density matrix, while the lower triangle subplots show the imaginary parts of the off-diagonal elements.}
\label{fig:caseBC_psd}
\end{figure*}

\begin{figure}
  \includegraphics[width=\columnwidth]{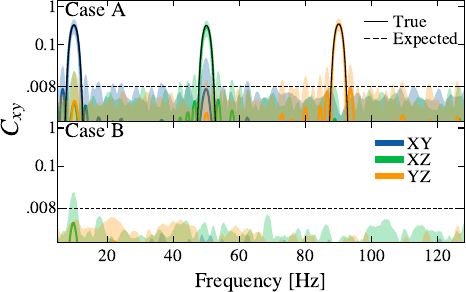}
  \caption{Squared coherence estimation for any pair of channels in cases A (top) and B (bottom). The solid curves represent the median estimated squared coherences, while the shaded areas indicate the corresponding 90\% credible interval.  The black solid line is the true squared coherence, and the black dashed line is the expected coherence from uncorrelated white noise data.}
  \label{fig:squared_coh}
\end{figure}

Figure~\ref{et_corr_basis_funs_vs_mle} displays the log maximum
likelihood estimates (MLE) across a range of $M$ for each case.
As the log MLE stabilizes around $M\sim400$, we set $M=450$.
Using the methodology discussed in Section~\ref{sec:method}, the PSDs for each case are estimated and plotted in Figure~\ref{fig:caseA_psd} and Figure~\ref{fig:caseBC_psd}. For each case, the diagonal subplots represent the spectral densities for X, Y and Z channels. Since the off-diagonal elements of the PSD matrix at each frequency are complex numbers, with the elements above the diagonal being the conjugates of those below, we plot the real part of the off-diagonal elements in the upper subplot and the imaginary part in the lower subplot.
Additionally, the squared coherences are plotted for the cases in Figure~\ref{fig:squared_coh}. The plot shows the squared coherences estimation between any pair of channels for case A and case B.

% SIDE BY SIDE VERSION
% \begin{figure*}[]
% \centering
% \begin{subfigure}{\columnwidth}
%   \centering
%   \includegraphics[width=1.05\columnwidth]{caseA_psd.pdf}
%   \caption{Case A PSD}
%   \label{fig:caseA_psd}
% \end{subfigure}
% \hfill
% \begin{subfigure}{\columnwidth}
%   \centering
%   \includegraphics[width=1.05\columnwidth]{caseBC_psd.pdf}
%   \caption{Case B and C PSDs}
%   \label{fig:caseBC_psd}
% \end{subfigure}
% \caption{Spectral density estimates for three cases of ET noise data. The plot shows the periodogram (gray), the PSD estimates (colored, with 90\% credible interval shaded) and the true PSD (black curve in the diagonals).}
% \label{fig:ET_PSDS}
% \end{figure*}

% For all of the three cases, it is evident that in the X channel, prominent peaks are observed at 10 Hz and 50 Hz. Similarly, for the Y channel, peaks are observed at 10 Hz and 90 Hz, while for the Z channel, peaks are observed at 50 Hz and 90 Hz. For case A, it is evident that there is significant coherence between channels X and Y at 10 Hz, between channels X and Z at 50 Hz, and between channels Y and Z at 90 Hz. For case B, although the ET noise data with uncorrelated noise is analysed, small fluctuations can still be observed in the off-diagonal subplots represent the real and imaginary part of the estimated spectral densities. The small correlations can also be observed from the  corresponding squared coherences. For case C, by treating the cross spectrum as zero, the estimated spectral densities for three channels are independent, this results in the square coherence is 0 as well.

In all three cases, prominent spectral peaks are observed at the specific frequencies in each channel, consistent with the injected Gaussian noise peaks.  
This demonstrates the ability of our proposed method to accurately capture and identify these peaks.
We note that our PSD estimates exhibit oscillatory features around the Gaussian peaks, this is most likely linked to the relatively loud and sharp Gaussian peak features. 
This highlights one of the potential weaknesses of the SGVB method: its potential difficulty in fitting very sharp features. 
For more realistic data sets including correlated noise from magnetic and/or Newtonian noise origin, this will most likely form less of an issue due to the smooth, power-law-like behavior of such noise sources \citet{Janssens_magnetic_noise,Janssens_newtonian_seismic,PhysRevD.109.102002,Janssens2023}. However, this could form an issue with actual detector data which is contaminated with a large amount of instrumental spectral noise lines. Future work will aim to investigate the capabilities of the SGVB method and its efficiency in deal with such complexity using real GW data and whether pre-cleaning steps should be undertaken. However, the dramatic speedup of PSD estimation using the SGVB remains an extremely potent benefit for next generation GW detectors. Without faster novel noise estimates techniques, such as the one proposed in this work, the resulting science will very likely be limited by available computation time.

In case A, where correlated noise is intentionally introduced, substantial coherence between channel pairs is observed at their shared peak frequencies (see the top panel of Figure~\ref{fig:squared_coh}). 
Specifically, high coherence is evident between channels X and Y at \SI{10}{Hz}, X and Z at \SI{50}{Hz}, and Y and Z at \SI{90}{Hz}. 
The median estimates for the squared coherence aligns with the true coherence, demonstrating that our estimate accurately reflects the underlying correlated noise structure. This alignment validates the effectiveness of our spectral analysis in capturing inter-channel relationships.

Case B, which considers ET noise data with nominally uncorrelated noise, reveals minor fluctuations in the off-diagonal subplots representing the real and imaginary parts of the estimated spectral densities, despite the absence of deliberately injected correlations. These fluctuations manifest as small coherence values (median $C_{xy} < 0.008$) that fluctuate around the true zero coherence in the corresponding squared coherence plots of Figure~\ref{fig:squared_coh} (visible as they are plotted on a log-scale). Notably, the median squared coherence fluctuations are lower than those expected from uncorrelated white noise data, which scale as approximately the inverse of the number of data segments used to average over. In this case, i.e.  $1/N_b = 0.008$. The squared coherence estimates based on the SGVB method are below this estimate and hence are consistent with zero coherence.

In case C, where the three channels are assumed independent, the model assumes zero cross spectrum (the XY, YZ, XZ components) in the PSD estimates.  Consequently, the squared coherence between any pair of channels is uniformly zero across all frequencies, which is consistent with theoretical expectations for uncorrelated channels, providing a baseline to assess the degree of correlation in the other cases.
Comparing case B and C highlights an important feature about the effectiveness of our SGVB model. 
As one would expect, a model with less complexity (case C) typically yields a more accurate estimate compared to a more complex model (case B). More concretely, the average root mean square deviation (RMSD)\footnote{The average RMSD is calculated as the mean of the square root of the mean squared differences between the estimated median PSD and the true PSD across the frequency band for X, Y and Z channels.} over the frequency band 5 to \SI{128}{Hz}  is $9.773 \times 10^{-50}$ for case B and $9.725 \times 10^{-50}$ for case C. The difference between both different scenario's is relatively small, case C's RMSD is about 0.5\% lower than that of case B, showcasing the huge potential and flexibility of the SGVB approach in estimating detector noise whether correlated noise is present or not.

\section{Conclusions}
\label{sec:Discussion}
Recent work has pointed out the importance of taking into account correlated interferometer network noise when estimating the parameters of gravitational wave signals observed by ground-based detectors such as ET or space-based detectors such as LISA~\cite{ET_design_report, Janssens2023, LISA_red_book}.
However, so far, no statistical methodology has been suggested to estimate the multivariate noise spectral density of a detector network.
In this study, we propose a computationally efficient variational Bayesian approach for estimating the spectral density of multivariate time series. 
It is important to note that this technique does not assume any restrictive parametric form of the spectral density, allowing it to adapt to any shape of the PSD.
To handle the large length of GW network measurements, we use a blocked Whittle likelihood. 
We provide a hyperparameter optimization method and guidelines to tune the settings of the method. 
Simulation studies presented for VAR(2) and VMA(1) demonstrate that our approach achieves accuracy comparable to MCMC while significantly reducing computation time, although with potentially narrower credible intervals.

The application of our method to simulated ET noise data revealed its ability to accurately identify and characterize spectral features across different noise scenarios.
In the correlated noise case, our approach successfully detected the injected Gaussian peaks and quantified the coherence between channels at specific frequencies.

These findings have significant implications for GW data analysis. The ability to efficiently and accurately estimate spectral densities and inter-channel correlations is crucial for optimizing detection algorithms and understanding noise characteristics in GW detectors. Our method's computational efficiency makes it particularly suitable for analyzing large-scale datasets expected from next-generation detectors (e.g. ET and LISA). The flexibility of our method, combined with its computational efficiency, makes it particularly suitable for analyzing large-scale datasets expected from next-generation GW detectors such as ET and LISA.

Future work could explore the application of this method to real GW detector data and its extension to simultaneously analyzing a stochastic GW background signal. Jointly analyzing the six channels of two detector networks such as TianQin and LISA could provide an more flexible alternative to cross-correlation detection  \cite{liang2024}.
Further investigations could examine the implications of these spectral characteristics on GW detection sensitivity and methods to mitigate correlated noise effects in ET or LISA. Extending the analysis to longer time series or different ET configurations could provide insights into the stability and generalizability of these spectral features.
Additionally, investigating the method's performance on GPUs could further validate its scalability, robustness and versatility.

In conclusion, the variational Bayes approach offers a promising tool for spectral density estimation in GW data analysis, combining accuracy with computational efficiency. This method has the potential to enhance our ability to characterize detector noise and ultimately improve GW detection capabilities.

%% TEMPORARILY COMMENTING OUT
%% WHILE WE CLEAN UP
\section*{Data and Software Availability}
The software developed for this research is available on GitHub under the MIT License~\citep{sgvb_psd_github}.
Documentation and examples of the software can be found at \url{https://nz-gravity.github.io/sgvb_psd/}. 
The raw data and data products for this work have been archived on Zenodo~\cite{sgvb_psd_zenodo}.

\begin{acknowledgments}
We thank Zhixiong Hu and Racquel Prado for making their code publicly available and for helpful discussions. AV,  JEL, NC, PMR and RM gratefully acknowledge support  by the Marsden Fund Council grant MFP-UOA2131 from New Zealand Government funding, managed by the Royal Society Te Apārangi. KJ was supported by FWO-Vlaanderen via grant number 11C5720N during part of this work.
We thank the New Zealand eScience Infrastructure
(NeSI) \url{https://www.nesi.org.nz} for the use of their high performance computing facilities and the Centre for eResearch at the University of Auckland for their technical
support.
\end{acknowledgments}

\vspace{4mm}
\noindent\textit{Software used:}
\python~\cite{python2020},
\tensorflowProb~\cite{tensorflow, tensorflowProb}
\numpy~\cite{numpy},
\scipy~\cite{scipy},
\pandas~\cite{pandas},
\matplotlib~\cite{matplotlib},
\jupyterbook~\cite{jupyterbook}.

\bibliography{reference}

\end{document}